\documentclass[superscriptaddress,twocolumn,showpacs,preprintnumbers,amsmath,amssymb,prl,10pt,aps]{revtex4-1}
\usepackage{graphicx}
\usepackage{dcolumn}
\usepackage{bm}
\usepackage{subfigure}
\usepackage{multirow}
\usepackage{soul}
\usepackage{float}
\usepackage[normalem]{ulem}
\usepackage[usenames,dvipsnames]{xcolor}
\usepackage{bbold}
\usepackage{hyperref} 

\hypersetup{backref,pdfpagemode=FullScreen,colorlinks=true,linkcolor=NavyBlue,citecolor=BrickRed}

\begin{document}

\title{Tunable Electronic Correlations in 135-Kagome Metals}

\author{Matteo Crispino}
\affiliation{Institut für Theoretische Physik und Astrophysik and Würzburg-Dresden Cluster of Excellence ct.qmat, Universität Würzburg, 97074 Würzburg, Germany}
\author{Niklas Witt}
\affiliation{Institut für Theoretische Physik und Astrophysik and Würzburg-Dresden Cluster of Excellence ct.qmat, Universität Würzburg, 97074 Würzburg, Germany}
\author{Stefan Enzner}
\affiliation{Institut für Theoretische Physik und Astrophysik and Würzburg-Dresden Cluster of Excellence ct.qmat, Universität Würzburg, 97074 Würzburg, Germany}
\author{Tommaso Gorni}
\affiliation{CINECA, via Magnanelli 2, Casalecchio di Reno (BO) 40033, Italy}
\author{Luca de' Medici}
\affiliation{LPEM, ESPCI Paris, PSL Research University, CNRS, Sorbonne Universit\'e, 75005 Paris France}
\author{Domenico Di Sante}
\affiliation{Department of Physics and Astronomy, Alma Mater Studiorum, University of Bologna, 40127 Bologna, Italy}
\author{Giorgio Sangiovanni}
\affiliation{Institut für Theoretische Physik und Astrophysik and Würzburg-Dresden Cluster of Excellence ct.qmat, Universität Würzburg, 97074 Würzburg, Germany}

\date{\today}
\begin{abstract}

Kagome metals exhibit rich correlated-electron physics, yet a systematic understanding of the degree of correlation across transition-metal species remains elusive. Using density-functional theory plus multi-orbital slave-spin mean-field theory, we investigate electronic correlations in the Ti-, V-, and Cr-based 135 compounds with Sb and Bi pnictogens. 
We find that the significantly stronger degree of correlation of the Cr-based materials compared to Ti and V can only be explained through the synergy of two effects: the larger electron filling of the $d$-shell and the reduced characteristic kinetic energy. We put forward that the substitution of Sb with Bi strengthens correlations in all compounds and make the prediction that the-yet-to-be-synthesized CsCr$_3$Bi$_5$ must be the most strongly correlated member of the entire family. These findings provide a quantitative, band-structure-based framework for understanding and predicting correlation strength in Kagome metals.

\end{abstract}

\pacs{}

\maketitle

\emph{Introduction --} Kagome lattice materials combine geometric frustration, topological band structures, and strong electronic correlations, creating a fertile ground for exotic quantum phenomena~\cite{di2025kagome,wilson2024v3sb5,wang2023quantum,Jiang2022review,neupert2022charge}. The AV$_3$Sb$_5$ family (A = K, Rb, Cs alkali atom)~\cite{Ortiz2019discovery,ortiz2021superconductivity,ortiz2020cs} has emerged as a paradigmatic system, with members exhibiting charge density waves (CDW), superconductivity, and signatures of time-reversal symmetry breaking~\cite{jiang2021unconventional,mielke2022time,Zhao2021cascade,Nakayama2021gap,Hu2022vanhove}. Angle-resolved photoemission spectroscopy (ARPES) reveals multiple van Hove singularities near the Fermi level and orbital-selective band renormalization, even though the overall agreement with first-principles calculations points towards moderate electronic correlations~\cite{kang2022twofold,Nakayama2022carrier,Hu2023landscape,Zhong2024arpes}. 

Recent synthesis of Cr-based analogs has revealed enhanced correlation signatures~\cite{sangiovanni2024superconductor,liu2024superconductivity}, including incipient flat bands and increased spectral incoherence~\cite{li2025electron,peng2026flat,Xie_Cr-135-Qimiao,wang2025heavy}, while Ti-substitution studies show systematic evolution of electronic structure~\cite{liu2023doping}, the appearance of superconductivity and high-temperature nematic order~\cite{yang2024superconductivity,li2023electronic,hu2023non,jiang2023flat,bigi2024pomeranchuk}, as well as anomalous spin-optical response~\cite{mazzola2025anomalous}. However, a quantitative framework for predicting correlation strength across the 135 family based on fundamental band structure properties remains lacking. Understanding this relationship is crucial for designing strongly correlated topological kagome materials and elucidating the microscopic origins of their correlation-driven physics.

The conventional single-orbital picture of strongly correlated systems suggests that the degree of correlation is fully encoded in the competition between kinetic energy (as measured, e.g., by the bandwidth $W$) and interaction strength ($U$)~\cite{Georges_DMFT-Rev,Held_DC,Qin_Hubbard}. 
Many-body effect typically depend also on the electron density. For example, in idealized single- and multi-orbital Hubbard models the strongest effects are normally reached at global half filling of the system.
Yet, kagome metals are characterized by a correlated manifold that is massively entangled to ligand bands. This makes the degree of correlation hard to quantify via a single scalar quantity. Not only the differences, but also the fluctuations between orbitals add complexity to the description.    
An estimator of correlations with a clear physical meaning has been proposed to be the proximity to half-filling of each individual orbital~\cite{deMedici_Giovannetti_Capone,Georges_deMedici_Mravlje_Rev,deMedici_Hund_corr}. 
However, a clear recipe for tuning correlations in 135 kagome metals is still missing.

In this Letter, we employ density-functional theory (DFT) plus multi-orbital many-body theory to understand correlation trends across Ti-, V-, and Cr-based 135 kagome metals with both Sb and Bi pnictogens. We demonstrate that the physics can only be explained by supplementing the distance from half filling of the individual orbital occupations with an indicator of the kinetic energy of each orbital. 
This understanding allows us to explain how Bi substitution to Sb enhances correlations and to predict CsCr$_3$Bi$_5$ as the most strongly correlated member of the 135 family.

\begin{figure*}[t]
    \centering
    \includegraphics[width=\linewidth]{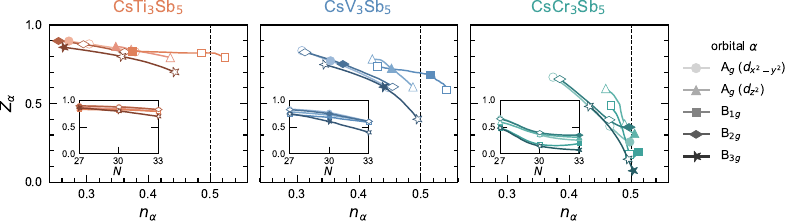}
    \caption{Orbital-resolved quasiparticle weights $Z_\alpha$ as a function of the orbital-resolved occupancy $n_\alpha$ of the transition metal. Orange, blue and green colors refer to CsTi$_3$Sb$_5$, CsV$_3$Sb$_5$, and CsCr$_3$Sb$_5$, respectively. For the different compounds, in going from left to right in each orbital, the symbols refer to a compound's total number of electrons ($N=$ 27, 30, and 33 for CsTi$_3$Sb$_5$, CsV$_3$Sb$_5$, and CsCr$_3$Sb$_5$, respectively). In each panel, the filled symbols mark the undoped occupancy of the corresponding compound, while the empty ones indicate the results of the simulations upon doping. Regardless of the $N$, CsCr$_3$Sb$_5$ shows the highest degree of renormalization. For each compound, $Z_\alpha$ as a function of $N$ are reported in the inset.}
    \label{fig:Sb-Z}
\end{figure*}
\emph{Model and method -- } 
For our theoretical study of the 135 kagome family we employ first-principles calculations based on DFT as implemented in \textsc{Quantum Espresso}~\cite{Giannozzi_Quantum-Espresso,Giannozzi_Quantum-Espresso-Advanced}. All the details can be found in the Supplemental Material~\cite{Crispino_Kagome-Supp}.  
We construct an effective Wannier function-based tight-binding model of 33 orbitals~\cite{marrazzo2024wannier}, by considering $3 \times 1$ (M-\emph{s}) + $3 \times 5$ (M-\emph{d}) + $5 \times 3$ (Pn-\emph{p}) orbitals, where M=Ti, V, Cr, and Pn=Sb, Bi. From this, we determine the hopping amplitudes and the local energies of the non-interacting model. Note that the isovalent substitution of Sb with Bi does not change the total number of electrons ($N$) which is 27 for Ti-based compounds, 30 for CsV$_3$Sb$_5$, and 33 for Cr-based compounds. 

Interactions are considered for the five $d$-orbitals of the transition metals M, and they are described by the multi-orbital Hubbard model~\cite{Hubbard_Hubbard_Ham, Kanamori_Kanamori_Ham}. All the other orbitals, i.e., the $5 \times 3$ $p$-orbitals of the ligands and the $3 \times 1$ $s$-orbital of the transition metals, form a non interacting manifold to which the correlated $d$-orbitals of M hybridize. The interacting problem is solved via slave-spin mean-field (SSMF) approach ~\cite{deMedici_SSMF_OSMT,Yu_Si_U(1)_Slave_Spin, Crispino_SSMF_AFM}, capable of capturing the orbital-dependent band renormalization due to electronic interaction. We assume a purely-local density-density form of the interaction, which gives results consistent with experimental evidences when used within DFT+SSMF~\cite{Hardy_BaFe2As2_SS,Crispino_d-electron-HF,Crispino_MnPns}. In this context, the strength of local-many body correlations is encoded in the quasiparticle weights, i.e., inverse mass enhancements, $Z_\alpha$ where $\alpha=$ $A_{g}(d_{x^2-y^2})$, $A_{g}(d_{z^2})$, $B_{1g}$, $B_{2g}$, $B_{3g}$ indicates $d$-orbitals of M transforming under the irreducible representations of the site symmetry group $D_{2h}$ for the $3f$ Wyckoff positions of space group 191 (see Supplemental Material~\cite{Crispino_Kagome-Supp}). In other words, the local reference system for each kagome site is chosen in accordance with
the principal axes of the site symmetry group~\cite{wu2021nature,zeng2025electronic,Xie_Cr-135-Qimiao}. Generally speaking, $Z_\alpha$ decreases from unity (non-interacting case) to zero marking a metal-to-insulator Mott transition~\cite{deMedici_SSMF_OSMT, deMedici_Hassan_OSMT}.

All calculations are performed at zero temperature and for the on-site interaction $U=6$ eV and a fixed Hund's coupling $J=0.18\,U$, in agreement with previous calculations on V-based compounds~\cite{Di-Sante_Kagome-Scaling,Jeong_cRPA-CsVSb}\footnote{We underline that in the slave-spin mean field an increase in the ratio $J/U$ improves the agreement of the results with more accurate methods such as dynamical-mean field theory. More details can be found in Ref.~\cite{Crispino_PhD}. From Ref.~\cite{Jeong_cRPA-CsVSb} follows $J\approx 0.13\,U$ corresponding to $J=0.18\,U$ in the slave-spin formalism.}. We use a fully-localized limit prescription~\cite{Czyzyk_FLL,Anisimov_Czyzyk_U_J_as_CS,Anisimov_LDA_U} to deal with the double counting of electronic interaction, which has proven to be a suitable choice within DFT+SSMF~\cite{Crispino_PhD}. 

\emph{Correlations in Sb-based kagome metals -- } Transition metal substitution in CsM$_3$Sb$_5$ (M = Ti, V, Cr), changes the total number of electrons. In order to compare how correlations vary among the members of the family upon chemical substitution, we perform calculations for $N$ ranging from 27 (pristine filling of CsTi$_3$Sb$_5$) to 33 (pristine filling of CsCr$_3$Sb$_5$) for all the compounds. The resulting quasiparticles weights $Z_{\alpha}$, shown in the inset of Fig.~\ref{fig:Sb-Z} as a function of $N$, reveal a clear enhancement of correlations upon doping towards $N=33$. This effect is significantly stronger in CsCr$_3$Sb$_5$ than in its Ti- and V-based counterparts.
In materials in which $d$-bands are weakly hybridized to a ligand $p$-manifold, such behavior can be attributed to the global proximity to half-filling~\cite{deMedici_Integer-Filling}, corresponding here to 33 electrons in 33 bands~\cite{Crispino_Kagome-Supp}.
In kagome metals, whose intertwined band structure of correlated and non-interacting bands reacts non-trivially to doping, the relationship between electron count and correlations is more subtle. An ionic calculation based on oxidation numbers would foresee a regime far from half filling, with $1.3$ and $0.3$ electrons globally occupying the $d$-shells for CsCr$_3$Sb$_5$ and CsV$_3$Sb$_5$, respectively, and a completely empty shell for CsTi$_3$Sb$_5$. This is inconsistent with the observed correlation trends.
In the main panels of Fig.~\ref{fig:Sb-Z}, we therefore plot $Z_\alpha$ as a function of the orbital-resolved occupancy $n_\alpha$ of the transition metal. Within each panel, filled symbols correspond to the pristine compounds, while empty symbols denote doped configurations.

At $N=27$ (leftmost points in each main panel of Fig.~\ref{fig:Sb-Z}), the Ti-based and V-based compounds are weakly-correlated metals ($0.83$ $\lesssim$ $Z_\alpha$ $\lesssim$ $0.9$ and $0.74$ $\lesssim$ $Z_\alpha$ $\lesssim$ $0.84$, respectively for CsTi$_3$Sb$_5$ and CsV$_3$Sb$_5$), with orbital occupancies between $\simeq 0.25$ and $\simeq 0.31$. Conversely, the less correlated orbitals of  CsCr$_3$Sb$_5$ ($A_g$ and $B_{2g}$) show a quasiparticle renormalization of $Z_\alpha\simeq 0.67$ and an orbital occupancy closer to half filling.
Artificially doping CsTi$_3$Sb$_5$ and CsV$_3$Sb$_5$ towards the number of electrons of CsCr$_3$Sb$_5$ ($N=33$) correlates the compounds. This effect is more evident in CsV$_3$Sb$_5$ than CsTi$_3$Sb$_5$, which remains overall weakly correlated. 

On the contrary, CsCr$_3$Sb$_5$ displays a strongly-correlated regime at its natural filling $N=33$, with an heavy quasiparticle renormalization $Z_{B_{3g}}\simeq 0.07$ for the $B_{3g}$ orbital, whose $n_{\alpha} \simeq0.5 $.
Further, CsCr$_3$Sb$_5$ displays a particularly
pronounced orbital-selective behavior.
At $N=33$, for instance, the quasiparticle weight spans a wide range, from $Z \approx 0.07$ to $Z \approx 0.35$. This spread originates from the different degrees of hybridization of the $d$-orbitals with the $p$-electron manifold~\cite{Xie_Cr-135-Qimiao}.
For all compounds, the $B_{3g}$ orbital is the most strongly correlated one throughout the entire filling range and for CsCr$_3$Sb$_5$ this has the smallest $Z$. Taken together, these trends demonstrate that, independent of the total electron count, CsCr$_3$Sb$_5$ is systematically more correlated than its Ti- and V-based counterparts.

This enhanced correlation can be attributed to the fact that the Cr-based orbitals lie systematically closer to orbital-resolved half filling (see Fig.~\ref{fig:Sb-Z}), which naturally promotes an orbital-selective Mott tendency. Nevertheless, it is remarkable that, despite all compounds exhibit similar orbital-resolved fillings around $N = 33$, their correlation strengths differ substantially: CsTi$_3$Sb$_5$ remains only weakly correlated, CsV$_3$Sb$_5$ is moderately correlated, and CsCr$_3$Sb$_5$ is the only strongly correlated one when its $d$-shell is close to be half filled. This compound-dependent spread implies that a reliable description of the hierarchy of correlation strength in the family must take into account the filling of the $d$-shell as well as the band structure of the various materials.
In the next paragraph, we give an explanation to this synergistic cooperation between filling and electronic structure.

\emph{Kinetic energy as the key factor --} 
We compute the second centered moment of the non-interacting $d$-orbital-resolved density of states $\rho_{\alpha}(\omega)$: $ \sigma^2_\alpha =  \int \left( \omega - \langle E_\alpha \rangle \right)^2 \rho_{\alpha}(\omega) d\omega$. 
When referred to an individual $d$-orbital with index $\alpha$, its square root can be used to effectively measure the associated kinetic energy~\cite{bulla2000linearized,Xie_Cr-135-Qimiao,Rohringer_Hubbard-3D} (see also Supplemental Material~\cite{Crispino_Kagome-Supp}). In spite of the complexity of the band structure of these materials comprising several overlapping bands and characterized by a massive entanglement between $d$- and $p$-manifolds, $\sigma_\alpha$ 
provides us with the relevant additional information to describe correlation effects in these compounds.
In fact, it allows for a comparison of the different materials at the same filling or of the different orbitals within the same compounds: the orbital with the smallest $\sigma_\alpha$ can indeed be reliably identified as the most correlated one. 

\begin{figure}
    \centering
    \includegraphics[width=\columnwidth]{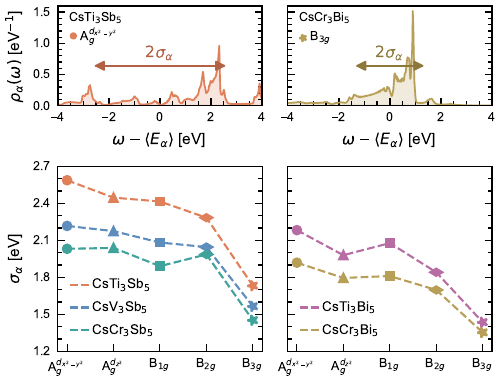}
    \caption{Top panels: projected density of states for the less correlated orbital ($A_g^{d_{x^2-y^2}}$ of CsTi$_3$Sb$_5$, on the left) and the most correlated orbital ($B_{3g}$ of CsCr$_3$Bi$_5$, on the right) of the series~\cite{Crispino_Kagome-Supp}. Bottom panels: Orbital-resolved standard deviation $\sigma_\alpha$ as determined by the square of the second centered moment of the density of states for CsM$_3$Sb$_5$ with M=Ti, V, Cr (left panel), and CsM$_3$Bi$_5$ with M=Ti, Cr (right panel).}
    \label{fig:Kagome-Ekin}
\end{figure}

In Fig.~\ref{fig:Kagome-Ekin} we show how there is a well-defined hierarchy among the Sb-based compounds: 
$\sigma^{\text{Cr}} < \sigma^{\text{V}} < \sigma^{\text{Ti}}$,
consistently across all orbitals.
In particular, the $B_{3g}$ orbital systematically exhibits the smallest {$\sigma$}.
This ranks CsCr$_3$Sb$_5$ as the member of this family with the closest-to-half-filling $d$-shell and with the highest $U$-over-kinetic energy ratio providing a robust explanation of the results shown in Fig.~\ref{fig:Sb-Z}. 
The ordering of the compounds at a given distance from the half-filled $d$-shell follows from this single number carrying information on the distribution of the spiky $d$-manifold as well as its intricate hybridization with the ligand-bands.
This establishes that the strongly correlated nature of CsCr$_3$Sb$_5$ originates from its electronic occupation in combination with a smaller spread of the non-interacting density of states of its $d$-orbitals. 
At the same time, the compound with the largest $\sigma$, namely CsTi$_3$Sb$_5$, remains weakly correlated across the entire filling range, even if we push its filling up to the same number of electrons of the Cr-compound.

\begin{figure}
    \centering
    \includegraphics[width=\columnwidth]{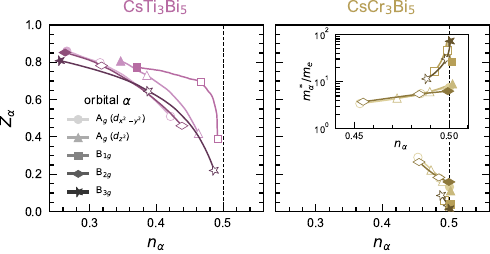}
    \caption{Orbital-resolved quasiparticle weights as a function of the orbital-resolved occupancy of the transition metal for the Bi-based kagome metals. Magenta and gold colors represent CsTi$_3$Bi$_5$ and CsCr$_3$Bi$_5$ respectively. Simulations' results at pristine filling, corresponding to $N=$ 27 for CsTi$_3$Bi$_5$ and to 33 for CsCr$_3$Bi$_5$, are marked by the filled symbols. Empty symbols refer to the results of simulations upon doping.}
    \label{fig:Bi-Z}
\end{figure}

\emph{Bi substitution effects --} Since CsTi$_3$Sb$_5$ has not been experimentally realized, a natural question concerns how these trends compare with the recently synthesized CsTi$_3$Bi$_5$~\cite{yang2024superconductivity}. The isovalent substitution of Bi for Sb preserves the electron count but increases the lattice parameters relative to CsV$_3$Sb$_5$ and CsCr$_3$Sb$_5$, thereby modifying the band structure. The resulting orbital-resolved quasiparticle weights, shown in the left panel of Fig.~\ref{fig:Bi-Z}, indicate that at $N=27$ the system remains a weakly correlated metal, with $0.77 \lesssim Z_\alpha \lesssim 0.86$, slightly more correlated than the Sb-based parent compound and comparable to CsV$_3$Sb$_5$. Remarkably, however, upon overdoping toward $N=33$, CsTi$_3$Bi$_5$ exhibits a dramatic enhancement of correlations, entering a regime comparable to CsCr$_3$Sb$_5$. As in the Sb-based series, the $B_{3g}$ orbital is the most strongly renormalized, reaching $Z_{B_{3g}}(N=33) \simeq 0.22$, which overcomes both its Sb analogue and CsV$_3$Sb$_5$. The remaining orbitals show quasiparticle weights in the range $0.39 \le Z_\alpha \le 0.51$, highlighting pronounced orbital selectivity, consistent with the behavior of CsCr$_3$Sb$_5$, but in stark contrast to the overdoped parent compound CsTi$_3$Sb$_5$ (leftmost panel of Fig.~\ref{fig:Sb-Z}).

\begin{figure}
    \centering
    \includegraphics[width=\columnwidth]{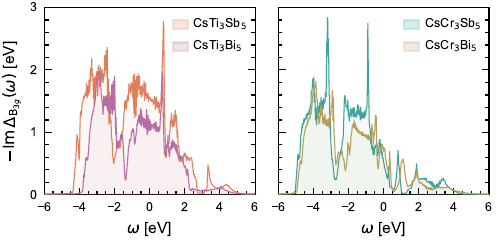}
    \caption{Imaginary part of the hybridization function $\Delta(\omega)$, specific to the most correlated $B_{3g}$ orbital, for the Sb- and Bi-derived compounds.}
    \label{fig:Delta}
\end{figure}

Interestingly, the unexpected strengthening of correlations in CsTi$_3$Bi$_5$ can be understood within the same framework used for the Sb-based compounds. As shown in the right panel of Fig.~\ref{fig:Kagome-Ekin}, its kinetic energy scale is strongly reduced, essentially matching that of CsCr$_3$Sb$_5$. The Bi-for-Sb substitution pushes the orbital occupancies closer to their respective half fillings as evident from Fig.~\ref{fig:Bi-Z}, and therefore enhances localization on the kagome sites. In this respect, the left panel of Fig.~\ref{fig:Delta} displays the imaginary part of the orbital-resolved hybridization function $\Delta_{B_{3g}}(\omega)$ for CsTi$_3$Sb$_5$ and CsTi$_3$Bi$_5$ (see Supplemental Material for details~\cite{Crispino_Kagome-Supp}). Upon pnictogen substitution, the overall spectral weight of the hybridization function is visibly reduced across the entire frequency window. Since the integrated weight of $-\mathrm{Im}\,\Delta(\omega)$ quantifies the effective coupling of the correlated $d$-orbitals to the itinerant degrees of freedom, its suppression signals a diminished hybridization strength. This reduction is naturally interpreted as an enhancement of orbital localization~\cite{Mravlje_Sr2RuO4}, consistent with the narrowed bandwidth and lower kinetic-energy scale identified in our analysis. In this sense, the behavior of the hybridization function provides a further independent, directly band structure–derived, confirmation that Bi substitution systematically strengthens electronic correlations in the 135 kagome family.

\emph{Prediction for CsCr$_3$Bi$_5$ --} Following this logic, CsCr$_3$Bi$_5$, combining the most correlated transition metal (Cr) with the correlation-enhancing pnictogen (Bi), should exhibit the strongest correlations. To address this, we simulate the not-yet-synthesized CsCr$_3$Bi$_5$. In the absence of experimental structural information about CsCr$_3$Bi$_5$, we adopt the same lattice constants of the Sb analogue~\cite{Liu_Cr-135-Pressure} and relax the internal atomic positions. The effect of this chemical substitution on the correlation strength is shown in the right panel of Fig.~\ref{fig:Bi-Z}: the minimal change of replacing Sb with the isovalent but heavier Bi drives a dramatic enhancement of correlations. CsCr$_3$Bi$_5$ becomes significantly more correlated than CsCr$_3$Sb$_5$ itself. In the underdoped regime, the quasiparticle weights already fall within $0.06 \lesssim Z_\alpha \lesssim 0.29$, markedly lower than in any of the Sb-compounds. At the pristine filling $N=33$, all orbitals shift even closer to their individual half-filled limits, yielding $0.01 \lesssim Z_\alpha \lesssim 0.16$. This places the compound on the verge of a Mott-localized regime (orbital-resolved mass renormalization $m^*_\alpha/m_e$ close to divergence, inset of Fig.~\ref{fig:Bi-Z}), with the $B_{1g}$ and $B_{3g}$ orbitals nearly Mott insulating and the remaining orbitals approaching the localization threshold.

Consistently, the kinetic energy scale of CsCr$_3$Bi$_5$, shown as the gold curve in Fig.~\ref{fig:Kagome-Ekin}, is the lowest among all materials considered, lying well below that of CsCr$_3$Sb$_5$ and CsTi$_3$Bi$_5$. Furthermore, the spectral weight of the hybridization function follows a similar reduction, as shown in the right panel of Fig.~\ref{fig:Delta}. These results further confirm the robustness of our interpretation based on orbital filling and band flattening. 
We stress that our prediction does not depend on the specific choice of lattice parameters:
Modeling CsCr$_3$Bi$_5$ using the larger lattice constants reported for CsTi$_3$Bi$_5$~\cite{Werhahn_TiBi-parameters} rather than those of CsCr$_3$Sb$_5$, enhances correlations even further.
(see Supplemental Material~\cite{Crispino_Kagome-Supp}). 
These results hence show that the simple isovalent substitution of Bi for Sb produces a substantial enhancement of electronic correlations in CsCr$_3$Bi$_5$, indicating a clear path to follow in order for the kagome 135 family to reach the extreme correlated limit.

\emph{Conclusions --} We have established a quantitative framework for understanding electronic correlation trends in 135 kagome metals based on fundamental band structure properties. Using multi-orbital slave-spin theory, we demonstrate that Cr-based compounds are significantly more correlated than Ti and V analogs. 
The two relevant ingredients leading to this are the
proximity to orbital-resolved half filling 
combined with a reduced degree of kinetic energy. 
Bi substitution universally further strengthens correlations motivating the prediction that CsCr$_3$Bi$_5$ would be the most strongly correlated member of the 135 family.

Our findings provide practical design principles for correlated topological materials within but also outside the kagome-metal platform. 
The systematic correlation enhancement through chemical substitution of the pnictogen (Bi for Sb) or of the transition metal (Cr for V or Ti) offers clear experimental pathways. Another way of tuning correlations in CsCr$_3$Sb$_5$ via pressure has been explored by M. Chatzieleftheriou et al.~\cite{Chatzieleftheriou_Cr-Pressure}.
The prediction made here of strong correlations in CsCr$_3$Bi$_5$ motivate urgent synthesis efforts, as this material may host enhanced regions of unconventional superconductivity, exotic magnetic states and novel topological phases arising from the interplay of strong correlations and kagome band topology.

\emph{Acknowledgments --} M.C. and G.S. thank Maria Chatzieleftheriou, Jonas Profe, and Roser Valentí for useful discussion. We gratefully acknowledge the Gauss Centre for Supercomputing e.V. (https://www.gauss-centre.eu) for funding this project
by providing computing time on the GCS Supercomputer SuperMUC-NG at Leibniz Supercomputing Centre (https://www.lrz.de). M.C. acknowledges financial support by the Deutsche Forschungsgemeinschaft (DFG, German Research Foundation) under Germany's Excellence Strategy through the W\"urzburg-Dresden Cluster of Excellence on Complexity and Topology in Quantum Matter-ct.qmat (EXC 2147, project-id 390858490). G.S. acknowledges financial support through project P5 of the FOR 5249 [QUAST] by the DFG Nr. 449872909. N.W. acknowledges support from the DFG-funded SFB 1170 ToCoTronics (project No. 258499086), and DDS acknowledges the DFG through SFB 1170 ToCoTronics Mercator Fellowship.

\bibliography{main.bbl}

\end{document}


\title{Supplemental Material for Tunable Electronic Correlations in 135-Kagome Metals}


\author{Matteo Crispino}
\author{Niklas Witt}
\author{Stefan Enzner}
\author{Tommaso Gorni}
\author{Luca de'~Medici}
\author{Domenico di Sante}
\author{Giorgio Sangiovanni}

\maketitle

\setcounter{page}{\pagenumbaa}
\thispagestyle{plain}

\section{Kagome structure and density-functional theory details}

\begin{table}[h]
\centering
\caption{Structural parameters of the Kagome 135 materials. Columns list the in-plane lattice constant $a = |\mathbf{a}_1| = |\mathbf{a}_2|$, the out-of-plane lattice constant $c = |\mathbf{a}_3|$, and the distances $d(\mathrm{M},\mathrm{Pn}_i) = || \mathbf{r}_{\mathrm{M}} - \mathbf{r}_{\mathrm{Pn}_i}||$ between the transition metal (M) and the two symmetry-distinct pnictogen sites Pn$_{1,2}$.}
\begin{tabular}{|l|c|c|c|c|c}
\hline
Kagome metal & $a$ (\AA) & $c$ (\AA) & $d(\mathrm{M},\mathrm{Pn_1})$ (\AA) & $d(\mathrm{M},\mathrm{Pn_2})$ (\AA) \\
\hline
CsCr$_3$Bi$_5$     & 5.49 & 9.24 & 2.75 & 2.83 \\
CsTi$_3$Bi$_5$     & 5.79 & 9.21 & 2.89 & 2.98  \\
CsCr$_3$Sb$_5$     & 5.49 & 9.24 & 2.75 & 2.69  \\
CsV$_3$Sb$_5$      & 5.49 & 9.23 & 2.75 & 2.76  \\
CsTi$_3$Sb$_5$     & 5.49 & 9.24 & 2.75 & 2.87  \\
\hline
\end{tabular}
\label{tab:SM_Parameters}
\end{table}

\begin{figure}
    \centering
    \includegraphics[width=\columnwidth]{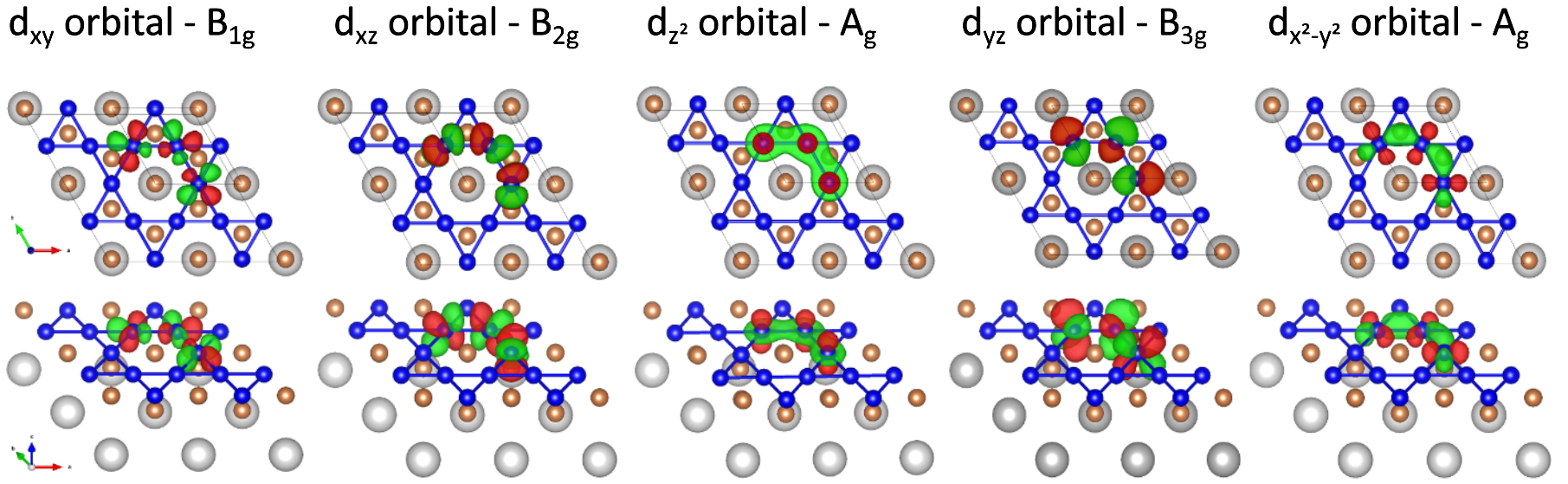}
    \caption{Symmetry-adapted Wannier wavefunctions for the five transition-metal $d$ orbitals.
From left to right: $d_{xy}$ ($B_{1g}$), $d_{xz}$ ($B_{2g}$), $d_{z^2}$ ($A_g$),
$d_{yz}$ ($B_{3g}$), and $d_{x^2-y^2}$ ($A_g$), labeled according to their
irreducible representations in the adopted point-group symmetry.
The Wannier functions are shown in the symmetry-adapted rotated basis used in the
construction of the low-energy Hamiltonian.
Upper panels show the top (in-plane) view, while lower panels display the
corresponding three-dimensional side view.
Isosurfaces of opposite phase are indicated by different colors, highlighting the
orbital orientation, nodal structure, and hybridization with surrounding ligand
states.}
    \label{fig:SM_Structure}
\end{figure}

For our density-functional theory (DFT) calculations, we apply the generalized gradient approximation~\cite{SM-Perdew_PBE} and use norm-conserving pseudopotentials~\cite{SM-Hamann_Pseudopotential} with a plane-wave cut-off of 84 Ry.
We sampled the Brillouin Zone on a $12\times12\times8$ $\Gamma$-centered mesh with a Gaussian smearing of 4\,mRy.
The lattice constants of the unit cells are chosen based on experimental parameters as available in Ref.~\cite{SM-Zhang_V-135-parameters} for CsV$_3$Sb$_5$, Ref.~\cite{SM-Liu_Cr-135-Pressure} for CsCr$_3$Sb$_5$, and Ref.~\cite{SM-Werhahn_TiBi-parameters} for CsTi$_3$Bi$_5$. Atomic positions relaxed until the forces are converged to below $10^{-7}$\,Ry/au. To simulate CsTi$_3$Sb$_5$ and CsCr$_3$Bi$_5$, whose lattice constants are hitherto unknown, we use the same values as of CsCr$_3$Sb$_5$. 

Fig.~\ref{fig:SM_Structure} shows the kagome structure and representation of M-$3d$ orbitals rotated to align with the local axes of each kagome site, such that they transform under the irreducible representations of the site symmetry group D$_{2h}$ of the 3f Wyckoff positions.

During relaxation, only the $z$ component of the pnictogen atom located below and above the center of the kagome triangles is allowed to vary, while all other atom positions are constrained by the D$_{6h}$ symmetry Wyckoff positions. Table~\ref{tab:SM_Parameters} summarizes the structural parameters, including the distances between the two symmetry-distinct pnictogen positions (Pn$_{1,2}$, 1=in-plane, 2=out-of-plane) and the transition metal (M). Our results show that substituting Sb to Bi increases the M--Pn$_2$ distance, whereas replacing Ti with V and Cr reduces the M--Pn$_2$ distance. These trends are chemically consistent with the larger 6$p$ orbitals of Bi compared to 5$p$ orbitals of Sb and the smaller $d$-orbital radius of Cr relative to V and Ti.

However, these simple structural trends alone do not account for the observed variation in correlation strength across the kagome material family (c.f.~Fig.~1 and 3 in the main text). This observation highlights that the band structure of the kagome 135 family arises from a complex interplay of orbital size, shape, and symmetry, which cannot be reduced to a single geometric parameter. Remarkably, despite this complexity, the orbital-resolved parameter of standard deviation $\sigma_\alpha$, which quantifies the effective kinetic energy by integrating over all structural details, successfully captures and correlates with the correlation strength observed in our SSMF calculations, as detailed in the main text.

\section{Slave-spin formalism for correlated-uncorrelated manifolds}

The results derived in this work for CsM$_3$Pn$_5$, with M = Ti, V, Cr, and Pn = Sb, Bi are obtained trough the multi-orbital Hubbard model~\cite{SM-Hubbard_Hubbard_Ham,SM-Kanamori_Kanamori_Ham} considering a correlated manifold ($d$ orbitals of the transition metal, labeled by the $\hat{d}$ operator) hybridizing with an uncorrelated one ($s$ orbitals of the transition metals and $p$ orbitals of the pnictogens, both labeled by the $\hat{p}$ operator). The Hamiltonian $\hat{H}$ is the sum of two contribution, namely a kinetic-energy term considering all the hoppings and local energies from DFT:

\begin{equation}\label{eqn:SM_Ham-Kin}
    \hat{H}_0=\sum_{iM\sigma}\epsilon_{iM}\hat{n}^d_{iM\sigma}+\sum'_{ijMM'\sigma}t^{MM'}_{ij}\hat{d}^\dagger_{iM\sigma}\hat{d}_{jM'\sigma}+\sum_{ijLL'\sigma}w^{MM'}_{ij}\hat{p}^\dagger_{iL\sigma}\hat{p}_{iL'\sigma}+\sum_{ijML\sigma}\left( v^{ML}_{ij}\hat{d}^\dagger_{iM\sigma}p_{jL\sigma} + \textrm{h.c.} \right)
\end{equation}

and a local (density-density) interaction term, involving only the $d$ orbitals of the transition metal:

\begin{equation}\label{eqn:SM_Ham-Int}
    \hat{H}_\mathrm{int}=U\sum_{iM}\hat{\Tilde{n}}^d_{iM\uparrow}\hat{\Tilde{n}}^d_{iM\downarrow}+(U-2J)\sum_{iM\neq M'}\hat{\Tilde{n}}^d_{iM\uparrow}\hat{\Tilde{n}}^d_{iM'\downarrow}+(U-3J)\sum_{iM<M'\sigma}\hat{\Tilde{n}}^d_{iM\sigma}\hat{\Tilde{n}}^d_{iM'\sigma}.
\end{equation}

In Eq.~(\ref{eqn:SM_Ham-Kin}) and Eq.~(\ref{eqn:SM_Ham-Int}), $M=m\nu$ is a composite index labeling the orbital and the intra-cell lattice site (Latin and Greek letters, respectively), $\sum'$ is limited to $iM\neq jM'$, and $\hat{\Tilde{n}}_{iM\sigma}=\hat{d}^\dagger_{iM\sigma}\hat{d}_{iM\sigma}-\frac{1}{2}$. Moreover, $U$ is the Coulomb repulsion and $J$ is the Hund's coupling

The problem is solved within the slave-spin mean-field (SSMF) ~\cite{SM-deMedici_SSMF_OSMT,SM-Crispino_SSMF_AFM} approach. The aim of the method is to simplify the many-body electronic interaction from a two body operator ($\hat{\Tilde{n}}_{iM\sigma}\hat{\Tilde{n}}_{iM'\sigma}$, with $\sigma = \uparrow, \downarrow$) to a one body term expressed in terms of an auxiliary $\frac{1}{2}-$spin. To each physical electronic degree of freedom is associated a pseudo-fermion degree of freedom (labeled as $f$ in what follows) and the z-component of a $\frac{1}{2}$ spin. Precisely, $S^z_{iM\sigma}=+\frac{1}{2}$ corresponds to an occupied state, and vice versa $S^z_{iM\sigma}=-\frac{1}{2}$ to an empty state. This procedure enlarges the Hilbert space, and it generates unphysical states that can be avoided by implementing the constraint:
\begin{equation}\label{eqn:SM_constraint}
    \hat{f}_{iM\sigma}^\dagger \hat{f}_{iM\sigma}=\hat{S}^z_{iM\sigma}+\frac{1}{2},\quad \forall iM\sigma
\end{equation}
which allows to count the total number of physical electrons with either the pseudo-fermion degree of freedom or with the slave-spin one.
From this procedure, the operatorial relation follows $\hat{d}_{iM\sigma}^{(\dagger)}$ $\to$ $    \hat{f}_{iM\sigma}^{(\dagger)}    \hat{O}_{iM\sigma}^{(\dagger)}$, where $\hat{O}_{iM\sigma}=\hat{S}_{iM\sigma}^{-}+c_{iM\sigma}\hat{S}^{+}_{iM\sigma}$ with $c_{iM\sigma}$ being an arbitrary gauge which we fix to retrieve the proper non-interacting limit~\cite{SM-deMedici_Hassan_OSMT}. Note that, since the interaction among electrons is considered for the $d$ manifold only, the slave-spin procedure does not affect the $\hat{p}_{iL\sigma}$ operators. \\
Under this relation, the Hamiltonian is rewritten as:
\begin{align}\label{eqn:SM_Ham-pd-SSMF}
\hat{H}&=\sum_{iM\sigma}\epsilon_{iM}\hat{n}^f_{iM\sigma}+\sum'_{ijMM'\sigma}t^{MM'}_{ij}\hat{f}^\dagger_{iM\sigma}\hat{f}_{jM'\sigma}\hat{O}^\dagger_{iM\sigma}\hat{O}_{jM'\sigma}+\sum_{ijLL'\sigma}w^{MM'}_{ij}\hat{p}^\dagger_{iL\sigma}\hat{p}_{iL'\sigma}\nonumber \\
&+\sum_{ijML\sigma}\left( v^{ML}_{ij}\hat{f}^\dagger_{iM\sigma}\hat{O}^\dagger_{iM\sigma}p_{jLss\sigma} + \textrm{h.c.}\right) + \hat{H}_{\textrm{int}}\left[ \hat{S}^z \right],
\end{align}
where, thanks to the slave-spin procedure, the interaction is now a single-body term in the auxiliary degrees of freedom ($\hat{S}^z_{iM\sigma}\hat{S}^z_{iM'\sigma}$, with $\sigma=\uparrow,\downarrow$).

The problem is solved within a variational approach, under two variational ans\"atze: first, we assume the factorization of the total wave function in terms of pseudo-fermion and slave-spin wave functions, i.e. $|\Psi_\mathrm{tot}\rangle=|\Psi_{fp}\rangle|\Phi_{S}\rangle$; second, we assume $|\Phi_S\rangle=\prod_i |\phi_S^i\rangle$, equivalent to a single-site mean field of the auxiliary particles. The details of the minimization procedure can be found in Refs.~\cite{SM-Crispino_SSMF_AFM, SM-Crispino_PhD}.
The outcome is an interacting system of (slave) spins: 
\begin{equation} \label{eqn:SM_Ham-Spin-pd}
\hat{H}_S=\sum_{iM\sigma}\left( h_{iM\sigma}\hat{O}^\dagger_{iM\sigma}+\textrm{h.c.}\right)+\sum_{iM\sigma}\lambda_{iM\sigma}\hat{S}^z_{iM\sigma}+\hat{H}_\textrm{int}\left[\hat{S}^z\right]
\end{equation}
which is self-consistently coupled to a system of free fermions
\begin{align} \label{eqn:SM_Ham-Ferm-pd}
\hat{H}_{fp}&=\sum'_{ijMM'\sigma}t^{MM'}_{ij} \sqrt{Z_{iM\sigma}} \sqrt{Z_{jM'\sigma}} \hat{f}^\dagger_{iM\sigma}\hat{f}_{jM'\sigma}+\sum_{ijML\sigma}\left[v^{ML}_{ij} \sqrt{Z_{iM\sigma}} \hat{f}_{iM\sigma}\hat{p}_{jL\sigma} + \textrm{h.c.} \right]\nonumber \\
&+\sum_{ijLL'\sigma}w^{LL'}_{ij}\hat{p}^\dagger_{iL\sigma}\hat{p}_{jL'\sigma}+\sum_{iM\sigma}\left(\epsilon_{iM}-\mu-\lambda_{iM\sigma}+\lambda^0_{iM\sigma}\right)\hat{n}^f_{iM\sigma}-\mu\sum_{iL\sigma}\hat{n}^p_{iL\sigma},
\end{align}
renormalized by the interaction trough the quasiparticle weights $Z_{iM\sigma} \equiv | \langle \hat{O}^{\dagger}_{iM\sigma} \rangle |^2$. In Eq.~(\ref{eqn:SM_Ham-Spin-pd}) and Eq.~(\ref{eqn:SM_Ham-Ferm-pd}), $\{ \lambda_{iM\sigma} \}$ is a set of Lagrange multiplier enforcing the slave-spin constraint Eq.~(\ref{eqn:SM_constraint}), and $\mu$ is the chemical potential we have added to consider doping. Moreover, in Eq.~(\ref{eqn:SM_Ham-Spin-pd})
\begin{equation}
\label{eqn:h_pd}
h_{iM\sigma}=\sum'_{jM'}t^{MM'}_{ij} \sqrt{Z_{jM'\sigma}} \langle \hat{f}^\dagger_{iM\sigma} \hat{f}_{iM'\sigma} \rangle +\sum_{jL}v^{ML}_{ij}\langle \hat{f}^\dagger_{iM\sigma}\hat{p}_{jL\sigma}\rangle.
\end{equation}
acts as a transverse filed for the system of interacting slave spins, and it is reminiscent of the pseudofermions' presence. Finally, the additional field $\lambda^0_{iM\sigma}$ arises from the appropriate choice of the gauge $c_{iM\sigma}$~\cite{SM-Crispino_SSMF_AFM}.

To properly consider the double counting of the interaction both at the DFT and many-body theory level, we implemented the fully-localized limit double-counting correction~\cite{SM-Czyzyk_FLL,SM-Anisimov_Czyzyk_U_J_as_CS,SM-Anisimov_LDA_U}. This results in an additional chemical-potential like term acting on the correlated subspace only:
\begin{equation}
    \mu_{\mathrm{DC}}^{FLL}=\left[ \left( 2M - 1\right) U - 5 \left( M-1\right) J \right] \bar{n} + \left( U - 3J \right) \left( \bar{n} -\frac{1}{2}\right)
\end{equation}
where $\bar{n}=\frac{1}{2M} \sum_{M\sigma} \langle \hat{n}^f_{M\sigma} \rangle$ . Note that $\langle \hat{n}^f_{M\sigma} \rangle$ is updated at each step of the minimization procedure to take into account the redistribution of charge among $d$ and $p$ manifolds~\cite{SM-Crispino_PhD}.

\section{Quasiparticle weights as a function of total filling}
\begin{figure} 
    \centering
    \includegraphics[width=\columnwidth]{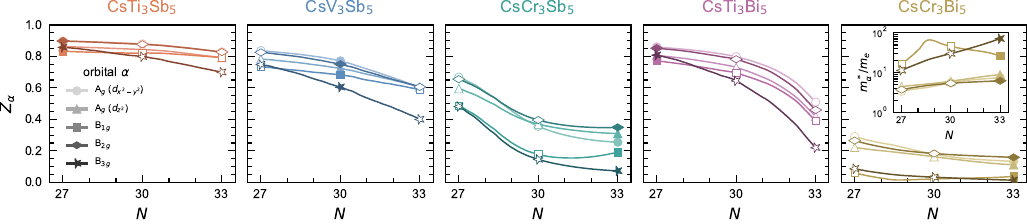}
    \caption{Orbital resolved quasiparticle weights $Z_\alpha$ as a function of total filling $N$ for CsM$_3$Pn$_5$, with M = Ti, V, Cr and Pn = Sb, Bi. DFT+SSMF simulations are done from the occupancy of pristine Ti-based compounds ($N=27$) to the one of pristine Cr-based compounds ($N=33$). For each compound, the filled symbols mark the corresponding undoped configuration. Empty points refer to pristine filling of the other compounds.}
    \label{fig:SM_Z-vs_N}
\end{figure}

In Fig.~\ref{fig:SM_Z-vs_N} we report the quasiparticle weights as a function of the total number of electrons $N$ for all members of the 135 kagome family. We simulate the range between the undoped occupancy of Ti-based compounds ($N=27$) to the one of Cr-based analogues ($N=33$). Among the CsM$_3$Sb$_5$, with M = Ti, V, and Cr, the Cr-based kagome metal is the most correlated compound at each filling, while CsTi$_3$Sb$_5$ remains weakly correlated irrespective of the total number of electrons. Bi substitution to Sb correlates the materials, making CsTi$_3$Bi$_5$ competing with CsCr$_3$Sb$_5$ and establishing CsCr$_3$Bi$_5$ as the most correlated member of the 135 kagome family.

\section{Second centered moment of the density of states and kinetic energy}
\begin{figure}
    \centering
    \includegraphics[width=0.95\columnwidth]{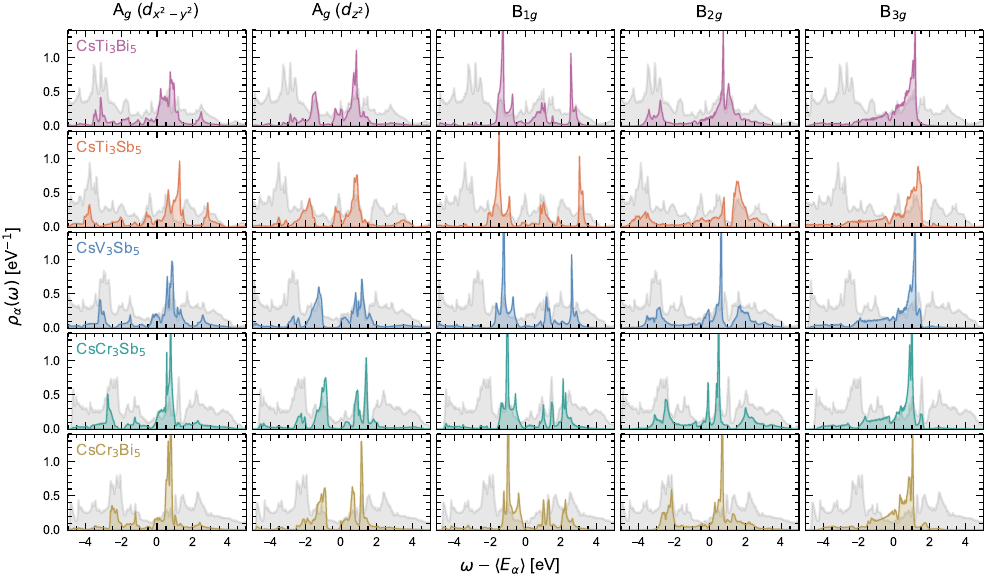}
    \caption{Orbital-resolved (projected) density of states for CsM$_3$Pn$_5$, with M = Ti, V, Cr and Pn = Sb,Bi. The gray background is the density of states of the uncorrelated manifold ($s$ orbitals of the transition metal and $p$ orbitals of the pnictogen). The smaller the variance of these distributions, the more the corresponding orbital is correlated (see main text).}
    \label{fig:SM_pDOS}
\end{figure}

We assume a general one-particle (tight-binding) Hamiltonian in an orthogonal (Wannier) basis $\lbrace | {\varphi_{\alpha} \rangle } \rbrace$ (compound index $\alpha=(\mathbf{R}_i, \sigma, m)$) which can be diagonalized in the Bloch basis $\lbrace | \psi_{n\bm{k}} \rangle \rbrace$ as:
\begin{align}
    \mathcal{H} = \sum_{\alpha\gamma} t_{\alpha\gamma} c^{\dagger}_\alpha c^{}_\gamma = \sum_{n\bm{k}} \varepsilon_{n\bm{k}} c^{\dagger}_{n\bm{k}}c^{}_{n\bm{k}}
\end{align}
via an unitary transformation $c_\alpha = \sum_{n\bm{k}} U_{\alpha n \bm{k}}^{} c_{n\bm{k}}$ (back: $c_{n\bm{k}} = \sum_{\alpha} U_{\alpha n \bm{k}}^* c_{\alpha}$) with $U_{\alpha n \bm{k}} = \langle {\varphi_\alpha}| {\psi_{n\bm{k}}} \rangle$. The total density of states for this Hamiltonian is given by (normalization implicitly included)
\begin{align}
    \rho(\omega) = \sum_{n \bm{k}} \delta(\omega - \varepsilon_{n \bm{k}})=\sum_{\alpha}\rho_\alpha (\omega)
    \label{eq:DOS_definition}
\end{align}
where $\rho_\alpha (\omega)$ is the orbital-resolved (projected) density of states:
\begin{align}
        \rho_\alpha(\omega) = \sum_{n \bm{k}} |\langle {\varphi_\alpha}|{\psi_{n\bm{k}}}\rangle|^2\; \delta(\omega - \varepsilon_{n\bm{k}})\;.
\end{align}

As proper probability distributions, we can define moments of the projected density of states. We are mainly interested in the first moment and second moment. The former describes the mean value:
\begin{align}
   \langle E_\alpha \rangle =\int_{-\infty}^{\infty}\! \rho_\alpha(\omega)\omega\,\mathrm{d}\omega\;.
\end{align}
For Wannier orbitals $\alpha = (\mathbf{R}_i,\sigma,m)$ this corresponds to the on-site energy (shifted by the chemical potential). In addition, we are interested in the second centered moment, describing the variance (standard deviation $\sigma$ squared):
\begin{align}
   \sigma^2_\alpha = \langle (E_\alpha-\langle E_\alpha\rangle)^2\rangle = \langle E_\alpha^2\rangle - \langle E_\alpha^{}\rangle^2=\int_{-\infty}^{\infty}\! \rho_\alpha(\omega)(\omega-\langle E_\alpha\rangle)^2\,\mathrm{d}\omega\;.
\end{align}
The results of the projected density of states for the compounds studied in the main text are reported in Fig.~\ref{fig:SM_pDOS}. The contributions from the individual M-3$d$ orbitals and the uncorrelated $p$ orbitals are shown, demonstrating the substantial overlap and strong entanglement of the $d$--$p$ electronic system.

\section{Hybridization Function}
The hybridization function $\Delta(\omega)$ of the non-interacting band structure is obtained from the local Green’s function projected onto the $d$-orbitals:
\begin{equation}
             \Delta(\omega) = (\omega + \mu)\mathbb{I}_{5\times5} - H_{d} - G_d^{-1}(\omega)\;.
\end{equation}

Here,  $\mu$ is the chemical potential, $H_d$ is the local part of the projected $d$-block of the non-interacting Hamiltonian, and $G_d(\omega)$ is the local Green’s function obtained from
\begin{equation}
             G(\omega) = \frac{1}{N_{\bm{k}}}\sum_{\bm{k}} [(\omega+\mu)\mathbb{I}_{31\times31} -H_{\bm{k}}]^{-1}    
\end{equation}

projected onto the $d$-orbitals of a single kagome site. The imaginary part, $-\mathrm{Im}\,\Delta(\omega)$, encodes the effective hybridization of the $d$-orbitals with the rest of the system, including both $d$--$d$ and $d$--$p$ contributions.

\section{Influence of lattice parameters on correlations in $\mathrm{Cs}\mathrm{Cr}_3\mathrm{Bi}_5$}
\begin{figure}
    \centering
    \includegraphics[width=0.8\columnwidth]{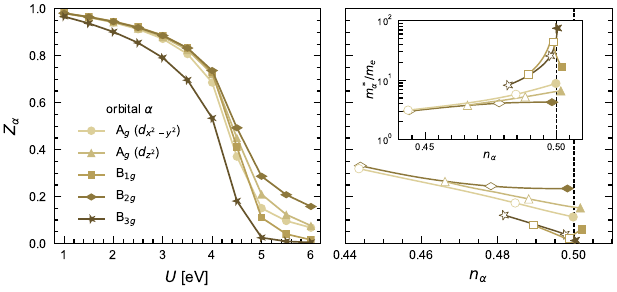}
    \caption{Left panel: Orbital-resolved quasiparticle weights $Z_\alpha$ as a function of the interaction. For $U > 5\,$eV, CsCr$_3$Bi$_5$ enters a strongly-correlated regime characterized by orbital selectivity in proximity to a Mott transition. Simulations are done at the pristine filling ($N=33$) and $J=0.18\,U$. Right panel: $Z_\alpha$ as a function of the orbital-resolved occupancy of Cr. Simulations at pristine filling ($N=33$) are marked by the filled symbols. Empty points refer to simulation' results upon doping. In the inset, the orbital-resolved mass renormalization is shown.}
    \label{fig:SM-large-CrBi}
\end{figure}

In the main text, we showed how isovalent substitution of Bi to Sb in CsCr$_3$Sb$_5$ strongly enhances the correlations, making CsCr$_3$Bi$_5$ the most correlated member of the family. Here, we provide further proof that this result is robust irrespective of the choice of the lattice parameters. We repeat the analysis performed in the main text for larger lattice constants, using the ones of CsTi$_3$Bi$_5$ as reported in Ref.~\cite{SM-Werhahn_TiBi-parameters}. The DFT+slave-spin simulations are shown in Fig.~\ref{fig:SM-large-CrBi}, where we report the quasiparticle weight $Z_\alpha$ as a function of interaction for $N=33$ (left panel) and as a function of the orbital-resolved occupancy in doping the system from $N=27$ to $N=33$ (right panel, calculations are done at $U=5.3\,$eV). In both calculations, $J=0.18\,U$. As shown, CsCr$_3$Bi$_5$ is strongly correlated across all dopings, and it follows the same conclusions of the main text, proving the robustness of our findings with respect to change in the lattice constants.

\clearpage
\bibliographystyle{unsrt}
\bibliography{supp-Kagome.bbl}